\documentclass[aps,prl,twocolumn,groupedaddress,showpacs]{revtex4}
\usepackage{natbib}
\usepackage{graphics}

\bibliography{12}

\begin{document}

\title{Topological spin transport of photons: the optical Magnus Effect and Berry Phase}
\author{K.Yu.Bliokh}
\affiliation{Institute of Radio Astronomy, 4 Krasnoznamyonnaya
st., Kharkov, 61002, Ukraine}
\author{Yu.P.Bliokh}
\affiliation{Department of Physics, Technion, Haifa, 32000,
Israel}

\begin{abstract}
The paper develops a modified geometrical optics (GO) of smoothly
inhomogeneous isotropic medium, which takes into account two
topological phenomena: Berry phase and the optical Magnus effect.
By using the analogy between a quasi-classical motion of a quantum
particle with a spin and GO of an electromagnetic wave in smoothly
inhomogeneous media, we have introduced the standard gauge
potential associated with the degeneracy in the wave momentum
space. This potential corresponds to the Dirac-monopole-like field
(Berry curvature), which causes the topological spin
(polarization) transport of photons. The deviations of waves of
right-hand and left-hand helicity occur in the opposite directions
and orthogonally to the principal direction of motion. This
produces a spin current directed across the principal motion. The
situation is similar to the anomalous Hall effect for electrons.
In addition, a simple scheme of the experiment allowing one to
observe the topological spin splitting of photons has been
suggested.
\end{abstract}

\pacs{41.20.Jb, 42.15.-i, 03.65.Vf, 05.60.-k}
\maketitle

%\date{\today}

The first consistent analysis of geometrical optics (GO)
asymptotic approximation for electromagnetic waves in a smoothly
inhomogeneous isotropic media was made in a work by Rytov
\cite{1}. There he studied the GO in zero approximation in small
parameter $k_{0}^{-1}=c/\omega$ ($\omega$ is the wave frequency);
also he gave a well-known law of rotation of electromagnetic wave
polarization plane, which corresponds to the first approximation
in $k_{0}^{-1}$. Geometrical attributes of this law were described
further in detail by Vladimirskiy \cite{2}. In the 1980s Berry
\cite{3} and his followers showed that geometrical phases are a
general fundamental attribute of dynamic systems, and the Rytov
Vladimirskiy polarization plane rotation is an example of Berry
phase (see \cite{4} and references therein).

Another fundamental phenomenon that is not contained in Rytov's GO
was discovered by Zel'dovich et al. in 1990 \cite{5}. It was named
``the optical Magnus effect'' and consists in that waves of right
and left circular polarization propagate in smoothly inhomogeneous
medium along different trajectories. Experimentally this
phenomenon was confirmed in waveguides in the mode limit. The
optical Magnus effect required a theoretical justification in the
framework of the GO, and phenomenological theory of this
phenomenon was proposed in \cite{6}, introducing some additional
correction terms in GO equations.

At the same time, recently a considerable attention has been
focused on a topological transport of quantum particles with a
spin (see \cite{7,8,9,10,11,12,13,14} and references there). This
problem is associated with the concepts of the spin splitting and
the spin pumping, and various questions of spintronics. The
topological spin transport and Berry phase are two manifestations
of the common phenomenon, namely, the initiation of a gauge
potential (connection) and a curvature in the space where the
particle transport occurs \cite{14}. This gauge potential and the
curvature affect both the particle's phase and its motion, giving
rise to an additional type of the transport. Since Berry phase
shows up in much the same way in nonrelativistic quantum particles
with a spin and photons (electromagnetic waves) \cite{4}, their
topological transport properties are bound to be similar. In this
case, the Berry phase corresponds to the Rytov law of polarization
evolution, while the topological spin transport corresponds to the
optical Magnus effect.

Below we construct the modified GO theory, which corresponds
exactly to the contemporary ideas of the topological spin
transport of particles. In so doing, we use the analogy between
the quasi-classical motion of particles with a spin and the wave
propagation in the GO approximation. Within the limits of this
correspondence, we should replace the Planck constant $\hbar$ with
the GO parameter $k_{0}^{-1}$, while discrete spin projections
correspond to discrete values of helicity, which can take the
values $\pm$ for photons \cite{15}. The latter corresponds to
right-hand and left-hand circular polarizations of waves. (It is
precisely the circularly polarized waves that are the eigenmodes
in the first approximation of GO \cite{15,16}.) The effects
considered (Berry phase and the Magnus optical effect) are
automatically accounted for when one introduces the gauge
potential that corresponds to the spin $1$ and the term degeneracy
(the intersection of the dispersion curves) at the origin in the
momentum space. It is well known \cite{3,4,9,10,13} that such
gauge potential gives rise to the force field like the Dirac
monopole field in the momentum space. In this case, two states of
the photon helicity (right-hand and left-hand circular
polarizations) correspond to two opposite charges in this field,
resulting in splitting of the wave trajectories. The obtained
results allow us to conclude that the optical Magnus effect is a
manifestation of the topological spin transport of photons.

As shown in Berry's pioneer work \cite{3}, the gauge potentials
and fields appear during the transport of the particle with a spin
and are due to the presence of the degeneracy points in the
particle's spectrum. In the generic case of the term intersection,
the universal expression can be written in adiabatic
(quasi-classical) approximation for a field tensor (Berry
curvature) $F_{ij}$:
\begin{equation}\label{eq1}
F_{ij}=m\varepsilon_{ijk}\frac{p_k}{p^3}~,~~\textrm{or}~~{\bf
F}=m\frac{{\bf p}}{p^3}~.
\end{equation}
Here ${\bf p}$ stands for a three-dimensional space, in which the
transport of the particle is considered (the generic case for the
points of degeneracy of Hermitian Hamiltonians corresponds
precisely to the three-dimensional space), the degeneracy point is
assumed to be in the origin of ${\bf p}$-space, $m$ is the
projection of the particle's spin (the helicity for relativistic
particles), and $\varepsilon_{ijk}$ is the unit antisymmetric
tensor. Since the ${\bf p}$-space is a three-dimensional one, we
write $F_{ij}$ in the form of a pseudovector ${\bf F}$,
Eq.(\ref{eq1}).

Field (curvature) (\ref{eq1}) corresponds to a certain gauge
potential (connection) ${\bf A}$ on the ${\bf p}$-space:
\begin{equation}\label{eq2}
  {\bf F}=\frac{\partial}{\partial{\bf p}}\times{\bf A}~,~~
  \textrm{or}~~
  F_{ij}=\frac{\partial A_j}{\partial p_i}-\frac{\partial A_i}{\partial p_j}~.
\end{equation}
The field ${\bf F}$ is invariant under gauge transformations of
potential, ${\bf A}\rightarrow {\bf A}+\partial\psi/\partial{\bf
p}$, and all basic laws for measurable quantities should be
expressed in terms of this field. If the coordinate or momentum
space acts as the ${\bf p}$-space, the potential enters into the
Hamiltonian and operates in perfect analogy to an electromagnetic
vector-potential in the appropriate space \cite{11,13}. Thus the
perfect analogy is seen between field (\ref{eq1}) and the magnetic
field of the Dirac monopole in the ${\bf p}$-space. Now the
effective 'spin charge' $m$ acts as the particle charge.

Let us apply these general statements to a photon or an
electromagnetic wave. Taking into account its dispersion
$\omega=\pm kc$ (or $E=\pm pc$, $E$ and $p$ are the photon energy
and momentum, respectively), we see that the point $p=k=0$ is the
point of the term intersection. Hence the gauge potential ${\bf
A}$ appears in the momentum space of the photon. This potential
corresponds to field (\ref{eq1}) with $m=\pm 1$ for right and left
circularly polarized photons.

Let us write the Hamiltonian of an electromagnetic wave in GO
approximation \cite{17}:
\begin{equation}\label{eq3}
  H({\bf p},{\bf r})=\frac{1}{2}\left[p^2-n^2 ({\bf r})\right]=0~.
\end{equation}
Here $n({\bf r})$ is the refractive index of the isotropic
smoothly inhomogeneous medium, while ${\bf p}=k_{0}^{-1}{\bf k}$
is the dimensionless momentum of the wave (${\bf k}$ stands for
the wavevector). Note that all calculations are conducted in the
first approximation in $k_{0}^{-1}$, that is, up to
$O\left(k_{0}^{-2}\right)$.

We must introduce into Hamiltonian (\ref{eq3}) the mentioned gauge
vector-potential ${\bf A}$ operating in the momentum ${\bf
p}$-space. As a result (compare with \cite{11,13}) we have for
circularly polarized waves:
\begin{equation}\label{eq4}
  H({\bf p},{\bf R})=\frac{1}{2}\left[p^2-
  n^2\left({\bf R}-k_{0}^{-1}{\bf A}({\bf p})\right)\right]=0~.
\end{equation}
Here we have switched from the usual coordinates ${\bf r}$ to the
\textit{generalized} coordinates ${\bf R}$,
\begin{equation}\label{eq5}
  {\bf r}={\bf R}-k_{0}^{-1}{\bf A}~,
\end{equation}
and have accounted for the correspondence $\hbar\leftrightarrow
k_{0}^{-1}$ between the quasi-classical quantum mechanics and GO.
By expanding Eq.(\ref{eq4}) in the Taylor series one can show that
it is physically equivalent to the Hamiltonian that has been
obtained in \cite{16}, and that follows immediately from Maxwell
equations. The problem in \cite{16} is considered in fact in the
generalized coordinates ${\bf R}$, since all properties are
derived there from the wave phases, which are determined just by
generalized coordinates (see Eq.(\ref{eq10})).

In the generalized coordinates the Hamilton ray equations for
Eq.(\ref{eq4}) have the canonical form (such equations were
studied in \cite{16}). However, the ray equations are more
conveniently considered in the ordinary coordinates ${\bf r}$,
Eq.(\ref{eq5}). By using the analogy with a motion of the charged
particle in a magnetic field, it is easily understood that the
presence of the vector-potential ${\bf A}$ in Eq.(\ref{eq4}) will
produce a force analogous to the Lorentz force with field
(\ref{eq1}), (\ref{eq2}) in a momentum space (see also
\cite{7,11,13}). As a result we have
\begin{equation}\label{eq7}
  {\bf \dot p}=-\frac{\partial H}{\partial{\bf r}}~,~~
  {\bf \dot r}=\frac{\partial H}{\partial{\bf p}}
  -k_{0}^{-1}\left({\bf F}\times{\bf \dot p}\right)~,
\end{equation}
where the dot signifies the differentiation with respect to the
parameter associated with the ray arc length. By substituting
Eq.(\ref{eq1}) with $m=\pm 1$ and Eqs. (\ref{eq3})--(\ref{eq5})
into Eq.(\ref{eq7}), we obtain
\begin{equation}\label{eq8}
  {\bf \dot p}=\frac{1}{2}\frac{\partial n^2}{\partial{\bf r}}~,~~
  {\bf \dot r}={\bf p}\mp
  k_{0}^{-1}\left(\frac{{\bf p}}{p^3}\times{\bf \dot p}\right)~.
\end{equation}

These are the desired equations, which describe the propagation
trajectory of electromagnetic waves (photons) in the first
approximation of GO. The first equation in (\ref{eq8}) and the
second equation without its last term, corresponds to the
well-known zero-order approximation of GO (see \cite{17}). The
last term in the right-hand side of the second equation in
(\ref{eq8}) determines the first-order correction in $k_{0}^{-1}$
and is of opposite signs for the waves of right-hand and left-hand
circular polarizations. It describes the Magnus optical effect or
the \textit{topological spin transport of photons}. We notice that
this term causes an additional displacement of photons of distinct
helicity in opposite directions normally to the ray. In other
words, the \textit{spin current} of photons across the principal
direction of their propagation arises. This situation is in
complete agreement with the currently considered topological spin
currents of quasi-classical particles \cite{7,8,9,10,11,12,13,14}.
Besides, the term obtained is quite similar to those describing
the anomalous Hall effect in a two-dimensional electron gas
\cite{8,9,10}. This analogy becomes all the more evident if one
notes that the anomalous Hall effect is related to the
spin-degeneration (in a momentum space) of the terms described by
the spin-orbital interaction. At the same time, the Magnus optical
effect also was associated with the spin-orbital interaction in
\cite{6}, and, as we have shown, is caused by the presence of the
point of degeneracy in a momentum space. Consequently, the Magnus
optical effect may be thought of as a manifestation of the
\textit{anomalous Hall effect of photons} \cite{18}.

The topological nature of trajectory splitting (\ref{eq8}) implies
that it is dictated exclusively by a gauge potential (connection)
and a field (curvature), which are connected with the degeneracy
in the momentum space. As a consequence, the magnitude of
splitting for the rays of left-hand and right-hand polarizations
is determined by a contour integral in the momentum space. Indeed,
by integrating the trajectory deviations, which are specified by
the last term of the second equation in (\ref{eq8}), we obtain
\begin{equation}\label{eq9}
\delta{\bf r}=\mp k_0^{-1}\intop_0^s{\frac{({\bf p}\times{\bf \dot
p})}{p^3}}ds=\mp k_0^{-1}\intop_L{\frac{({\bf p}\times d{\bf
p})}{p^3}}~,
\end{equation}
where $L$ is the contour in the ${\bf p}$-space along which the
system moves. Hence the optical Magnus effect is a topological
nonlocal effect as the Berry phase is.

The Rytov-Vladimirsky-Berry phase in the theory suggested
(Eqs.(\ref{eq4})--(\ref{eq8})) arises in perfect analogy to the
Dirac phase of a quantum particle in an electromagnetic potential.
Indeed, the wave phase (eikonal) in GO is to be expressed in terms
of \textit{generalized} coordinates, since the action is a
function of generalized coordinates. As a result, by constructing
the wave phase in terms of the generalized coordinates ${\bf R}$
and passing on to the usual coordinates ${\bf r}$, we have
\begin{equation}\label{eq10}
\varphi=-\omega t +k_0\intop_0^s{{\bf p}}d{\bf R}=-\omega
t+k_0\intop_0^s{{\bf p}}d{\bf r}-\intop_0^s{{\bf A}}d{\bf p}~.
\end{equation}
Here we have used Eq.(\ref{eq5}) and performed the integration by
parts; the term outside the integral is equal to zero because
${\bf Ap}=0$ for the gauge potential in question. The first two
terms and the third term in the right side of Eq.(\ref{eq10}) are
the ordinary dynamic and geometric phases, respectively. The
geometric Berry phase in Eq.(\ref{eq10}), being given by a contour
integral in the momentum ${\bf p}$-space, is equal for closed
contours to a flux of the field ${\bf F}$, Eq.(\ref{eq1}), through
the surface spanned on this contour. Clearly this phase is of
opposite signs for the waves of right-hand and left-hand circular
polarizations. In the general case of superposition of two
polarizations, the geometric phase leads to the Rytov law about
the rotation of the electromagnetic wave polarization plane
\cite{4}.

From the second equation in (\ref{eq8}) and equation (\ref{eq10})
the expressions for group and phase velocities of circularly
polarized waves follow immediately:
\begin{equation}\label{eq11}
{\bf v}_{g}\equiv -\frac{\partial H}{\partial {\bf k}}/
\frac{\partial H}{\partial\omega} =\frac{c}{n^2}{\bf\dot r}
=\frac{c}{n}\left[{\bf l}\mp k_{0}^{-1}\frac{({\bf p}\times{\bf
\dot p})}{p^3}\right]~,
\end{equation}
\begin{equation}\label{eq12}
{\bf v}_{ph}\equiv\frac{\partial\varphi/\partial
t}{|\partial\varphi/\partial{\bf
r}|^2}\frac{\partial\varphi}{\partial {\bf r}} = \frac{c}{n}{\bf
l}\left[1+k_{0}^{-1}\frac{{\bf A\dot p}}{p}\right]~.
\end{equation}
Here we have considered the medium as being dispersion-free
($n(\omega)=const$), took into account that $p=n$ (see
Eqs.(\ref{eq3}) and (\ref{eq4}), and kept in mind that the
geometric phase in Eq.(\ref{eq10}) is defined over the solutions
of zero-order approximation (for details see \cite{16}). It
follows from Eqs.(\ref{eq11}) and (\ref{eq12}) that the spin
transport of photons and the optical Magnus effect correspond to
the difference in \textit{group} velocities of the waves, whereas
the Berry phase corresponds to the difference in their
\textit{phase} velocities. The group velocities differ mainly in
\textit{direction} (the corrections are directed across the ray),
whereas the phase velocities differ in \textit{absolute value}.
The inequality of phase and group wave velocities results in a
weak \textit{anisotropy} of the locally isotropic medium
\cite{16}. Evidently the anisotropy results from the initiation of
the preferential direction, which is determined by the gradient of
the inhomogeneity ${\bf\dot p}\propto\partial n/\partial{\bf r}$
(see Eqs.(\ref{eq8})). Note also that the values of the phase
velocities are specified by the gauge potential ${\bf A}$ and
hence depend on the gauge transformations. This fact does not also
lead to contradictions, as the phase velocity here (in contrast to
the phase) is not a measurable value.

It should be noted that when substituting ${\bf\dot p}$ from the
first equation in (\ref{eq8}) into the second one, we obtain the
ray equations in the form suggested by Zel'dovich and Liberman in
\cite{6}. Meanwhile, there exist fundamental differences between
Ref. \cite{6} theory and the present work along with paper
\cite{16}. First of all, attention may be drawn to the difference
in the Hamiltonian for the waves in paper \cite{6} and in formula
(\ref{eq4}) of our work. It is evident that Ref. \cite{6} theory
involves many excessive variables, which complicate the problem.
In essence, the wave polarization was treated there as an
independent dynamical variable with a continuous spectrum of
values. At the same time, it is shown in \cite{16} that only the
right-hand and left-hand circularly polarized waves are the
independent eigen modes in the first approximation of GO. This is
consistent with the quantization of photon helicity \cite{15}. As
a consequence of this fact, an arbitrary wave of mixed
polarization splits into two waves of right-hand and left-hand
circular polarizations when propagating through an inhomogeneous
medium. This phenomenon of wave splitting into two independent
modes is absent in the paper \cite{6} (for example, according the
theory in \cite{6}, the trajectory of a linearly polarized wave
remains unchanged), but it is in complete agreement with the
contemporary conception of the topological spin transport of
quantum particles.

Two examples of spin (polarization) photon transport and splitting
are presented in Fig.1a and b. There we can see the trajectories
of right-hand and left-hand circularly polarized waves in circular
waveguides. Fig. 1a is associated with the optical Magnus effect
for the meridional ray in a circular waveguide (see \cite{5}),
whereas Fig.1b shows the splitting of the linearly polarized
finite ray into two rays of right-hand and left-hand polarizations
and their subsequent transport along the circular waveguide in
opposite directions (see detailed description of these two
examples in \cite{16}). The experiment presented in Fig.1b can be
performed in a cylindrical structure by exciting an oriented ray
associated with the mode of whispering gallery. This experiment
will enable one to observe for the first time the spin splitting
of photons.

%\begin{widetext}
\begin{figure}[t]
  \centering \scalebox{0.5}{\includegraphics{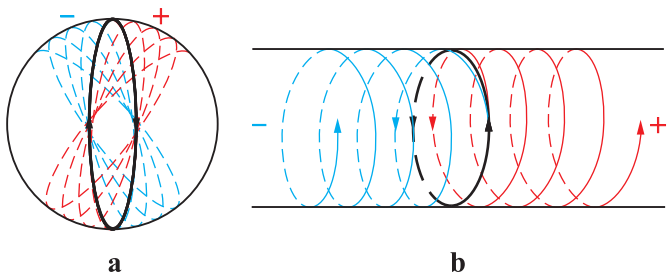}}
  \caption{The trajectories of right-hand (+) and left-hand (-) circularly
  polarized waves in circular waveguide. Fig.1a shows
  the shifts of the rays relative to the trajectory
  of the zero approximation (thick line) in the optical Magnus effect
  for the meridional ray \cite{5} (view from the waveguide end).
  Fig.1b shows the splitting of the linearly polarized finite ray
  into two circularly polarized eigen rays and their
  subsequent transport along the waveguide in opposite directions (angle view)}
  \label{}
\end{figure}
%\end{widetext}

Let us summarize the main outcomes of our work. Exclusively from
both the topological considerations and perfect analogy between GO
and a quasi-classical motion of quantum particles with a spin, we
have constructed the simple and clear scheme of the first GO
approximation in $k_{0}^{-1}$ for electromagnetic waves. By
introducing in a momentum space the standard gauge potential,
which is connected with the point of term degeneracy at the
origin, it is possible to describe such phenomena as the optical
Magnus effect \cite{5,6} and the Rytov-Vladimirsky-Berry geometric
phase \cite{1,2,3,4}. It becomes apparent that the optical Magnus
effect is nothing but the topological spin (polarization)
transport of photons, which is similar to the anomalous Hall
effect for electrons \cite{8,9,10}. It has been shown that the
spin photon transport and the optical Magnus effect correspond to
the difference in directions of the group velocities of the
oppositely polarized waves, whereas the Berry phase is associated
with the difference in absolute values of their phase velocities
\cite{16}. In the context of the approach suggested the theory is
simple and clear and is in perfect agreement with the previous
direct derivations \cite{16} following from Maxwell equations. In
addition, our theory describes a novel effect, namely, the
splitting of a ray of mixed polarization into two rays of
right-hand and left-hand circular polarizations. The simple scheme
of the experiment for observing this phenomenon has also been
suggested (Fig.1b and \cite{16}).


\begin{references}
\bibitem{1} S.M. Rytov, Dokl. Akad. Nauk. SSSR {\bf 18}, 263 (1938).

\bibitem{2} V.V. Vladimirskiy, Dokl. Akad. Nauk. SSSR {\bf 31}, 222 (1941).

\bibitem{3} M.V. Berry, Proc. R. Soc. A {\bf 392}, 45 (1984).

\bibitem{4} A. Shapere and F. Wilczek (ed.), {\it Geometric Phases in Physics}
(Singapore: World Scientific, 1989); S.I. Vinitskiy, V.L. Debrov,
V.M. Dubovik, B.L. Markovski, and Yu.P. Stepanovskiy, Uspekhi
Fizicheskih Nauk {\bf 160}, 6 (1990) [Sov. Phys. Usp. {\bf 33}, 6
(1990)].

\bibitem{5} A.V. Dooghin, N.D. Kundikova, V.S. Liberman, and B.Ya. Zel'dovich,
Phys. Rev. A, {\bf 45}, 8204 (1992); B.Ya. Zel'dovich and V.S.
Liberman, Kvant. Elektron. {\bf 17}, 493 (1990) [Sov. J. Quantum
Electron. {\bf 20}, 427 (1990)]; A.V. Dooghin, B.Ya. Zel'dovich,
N.D. Kudnikova, and V.S. Liberman, Pis'ma Zh. Eksp. Teor. Fiz.
{\bf 53}, 186 (1991) [Sov. Phys. JETP Letters {\bf 53}, 197
(1991)]; A.V. Dooghin, B.Ya. Zel'dovich, N.D. Kudnikova, and V.S.
Liberman, Zh. Eksp. Teor. Fiz. {\bf 100}, 1474 (1991) [Sov. Phys.
JETP {\bf 73}, 816 (1991)].

\bibitem{6} V.S. Liberman and B.Ya. Zel'dovich, Phys. Rev. A {\bf 46}, 5199
(1992).

\bibitem{7} G. Sundaram and Q. Niu, Phys. Rev. B {\bf 59}, 14915 (1999).

\bibitem{8} T. Jungwirth, Q. Niu, and A.H. MacDonald, Phys. Rev. Lett. {\bf 88}, 207208 (2002).

\bibitem{9} D. Culcer, A. MacDonald, and Q. Niu, Phys. Rev. B {\bf 68}, 045327 (2003).

\bibitem{10} A.M. Dudarev, R.B. Diener, I. Carusotto, and Q. Niu, cond-mat/0311356v1.

\bibitem{11} S. Murakami, N. Nagaosa, and S.-C. Zhang, Science {\bf 301}, 1348 (2003).

\bibitem{12} B.A. Bernevig, J.P. Hu, E. Mukamel, and S.-C. Zhang, cond-mat/0311024.

\bibitem{13} F. Zhou, cond-mat/0311612.

\bibitem{14} H.-Q. Zhou, S.Y. Cho, and R.H. McKenzie, Phys. Rev. Lett {\bf 91}, 186803 (2003).

\bibitem{15} V.B. Berestetskiy, E.M. Lifshits, and L.P.Pitaevskiy {\it Relativistic Quantum Theory, V.1}
(Nauka, Moscow, 1968).

\bibitem{16} K.Yu. Bliokh and Yu.P. Bliokh, physics/0402014.

\bibitem{17} Yu.A. Kravtsov and Yu.I. Orlov, {\it Geometrical Optics of Inhomogeneous Medium}
(Nauka, Moscow, 1980; Springer-Verlag, Berlin, 1990).

\bibitem{18} Note, that ordinary Hall effect for photons in magnetic field also was discovered in recent works
B.A. van Tiggelen, Phys. Rev. Lett. {\bf 75}, 422
(1995); G.L.J.A. Rikken and B.A. van Tiggelen, Phys. Rev. Lett.
{\bf 78}, 847 (1997); D. Lacoste, F. Donatini, S. Neveu, J.A.
Serughetti, and B.A. van Tiggelen, Phys. Rev. E {\bf 62}, 3934
(2000).

\end{references}
\end{document}